\documentclass{ifacconf}

\usepackage{natbib}            %
\usepackage{graphicx}          %
\RequirePackage[OT1]{fontenc}
\RequirePackage{amsmath,amssymb}
\numberwithin{equation}{section}
\newtheorem{lemma}{Lemma}[section]
\newtheorem{remark}[lemma]{Remark}
\newtheorem{theorem}[lemma]{Theorem}
\newtheorem{corollary}[lemma]{Corollary}

\newtheorem{algor}[lemma]{Algorithm}
\newcommand{\B}[1]{{\bf #1}}

\newcommand{\R}[1]{{\rm #1}}
\newcommand{\mB}[1]{{\mathbb{#1}}}
\newcommand{\set}[2]{\left\{#1\,\left\vert\, #2\right.\right\}}

\begin{document}

\begin{frontmatter}

\title{Robust and Trend-following Kalman Smoothers using Student's t
%\thanksref{footnoteinfo}
} %

\author[First]{Aleksandr Aravkin}
\author[Second]{James V. Burke}
\author[Third]{Gianluigi Pillonetto}

\address[First]{Department of Earth and Ocean Sciences, University of British Columbia, Vancouver, Canada (e-mail: saravkin@eos.ubc.ca).}
\address[Second]{Department of Mathematics, University of Washington, Seattle, USA (e-mail: burke@math.washington.edu)}
\address[Third]{Department of Information Engineering, University of Padova, Padova, Italy (e-mail: giapi@dei.unipd.it)}


\begin{abstract}                          
We propose two nonlinear Kalman smoothers
that rely on Student's t distributions.
The {\it{T-Robust smoother}} 
finds the maximum {\it a posteriori}
likelihood (MAP) solution for Gaussian process noise and 
Student's t observation noise, and is extremely 
robust against outliers, outperforming the recently
proposed $\ell_1$-Laplace smoother in extreme situations
(e.g. 50\% or more outliers). 
The second estimator, which we call the
{\it{T-Trend smoother}}, is 
able to follow sudden changes in the process model,
and is derived as a MAP solver
for a model with Student's t-process noise and Gaussian 
observation noise. We design specialized methods to 
solve both problems which exploit the special structure
of the Student's t-distribution, and provide a convergence theory.
Both smoothers can be implemented with only minor modifications
to an existing $L_2$ smoother implementation. 
Numerical results for linear and nonlinear models
illustrating both robust and fast tracking applications are presented. 
\end{abstract}

\end{frontmatter}

\section{Introduction}

The Kalman filter is an efficient 
recursive algorithm that estimates the state of 
a dynamic system from measurements
contaminated by Gaussian noise \citep{kalman}. 
Along with many variants and extensions, 
it has found use in a wide array of applications 
including navigation, medical technologies and
econometrics \citep{Chui2009,West1991}.
Many of these problems are nonlinear, and 
may require smoothing over past data in both 
online and offline applications to improve significantly
the estimation performance \citep{Gelb}.\\ 
In this paper, we focus on two important 
areas in Kalman smoothing: robustness with respect to 
outliers in measurement data, and 
improved tracking of quickly changing system dynamics. 
Robust smoothers have been a topic of significant
interest since the 1970's,
e.g. see \citep{Schick1994}. In order to design 
outlier robust smoothers, 
recent reformulations described in \citep{AravkinL12011, AravkinIFAC,Farahmand2011} use 
$L_1$, Huber or Vapnik loss functions in place of $L_2$ penalties. 
There have also been recent efforts to
design smoothers that are able to better track fast system
dynamics, e.g. jumps in the state values. 
For example, in \citep{Ohlsson2011}
the Laplace distribution is used in place of the Gaussian distribution 
to model transition noises.
This introduces an $L_1$ penalty on the
state evolution in time and can be interpreted
as a dynamic version of the well known LASSO procedure
\citep{Lasso1996}.\\ 
All of these smoothers can be derived from a statistical point of view,
using log-concave densities 
on process noise, measurement noise,
and prior information on the state.  Log-concave densities take the form 
\begin{equation}
\label{logconcave}
\B{p}(\cdot) \propto \exp(-\rho(\cdot)), \quad \rho \text{ convex}\;.
\end{equation}
Formulations using~\eqref{logconcave} are nearly ubiquitous, 
in part because they correspond to convex optimization problems
in the linear case. However, to effectively model a regime with large outliers
or sudden jumps in the state, we want to look beyond~\eqref{logconcave}
in order to allow heavy-tailed distributions. \\
A particularly convenient heavy-tailed modeling distribution that 
falls outside of \eqref{logconcave} is the Student's t-distribution. 
In the statistics literature, this distribution was successfully applied to a 
variety of robust inference applications \citep{Lange1989}, 
and is closely related to re-descending influence functions \citep{Hampel}.
In the context of Kalman filtering/smoothing, the idea of using 
Student's t-distributions to model both measurement error and 
innovations (process errors) was studied in \citep{Fahr1998}. \\
We propose new nonlinear Kalman smoothers which 
we call T-Robust and T-Trend.  
For the T-Robust smoother, we model the measurement
noise using the Student's t-distribution, extending the approach 
in \citep{AravkinL12011} to the Student's t.
As a result, errors or 
`outliers' in the measurements have even 
less effect on the smoothed estimate, and performs better than
\citep{AravkinL12011} for cases with high proportion of outliers 
(e.g. 50\% bad data).  
For the T-Trend smoother, we instead
model process noise using the Student's t-distribution, which allows
the smoother to track sudden changes in state. \\
Our work differs for \citep{Fahr1998} in two important respects. 
First, we include {\it nonlinear measurement and process models}
in our analysis. Second, we propose a novel optimization algorithm
to solve both T-Robust and T-Trend smoothing problems.
This algorithm differs significantly from the one proposed in \citep{Fahr1998}. 
In particular, rather than using the Fisher information matrix (i.e. 
full Hessian), or its expectation as a Hessian approximation (method
of Fisher's scoring) as suggested by \citep{Fahr1998}, 
we design a modified Gauss-Newton method which builds
information about the curvature of the Student's t-log likelihood
into the Hessian approximation, and provide a convergence theory. 
The fast smoothing procedures proposed here also 
allow the efficient estimation of hyperparameters such as degrees of 
freedom (e.g. cross-validation techniques using the EM algorithm are 
discussed in \citep{Fahr1998}).  \\ 
The paper is organized as follows. 
In section \ref{StudentT-KF}
we introduce the multivariate Student's t-distribution 
and define the models underlying the T-Robust and T-Trend smoothers.
In Section \ref{ML} the maximum likelihood objective for T-Robust, 
the quadratic approximation for this objective,
and the convex quadratic program to solve this 
approximate subproblem are reported. 
In Section \ref{T-Trend} the T-Trend smoother
is designed by modeling transitions using 
Student's t.
We describe our algorithm, provide a convergence
theory, and explain the differences with \citep{Fahr1998}
in \ref{Algorithm}.  
The T-Robust and T-Trend smoothers are tested
using simulated data for linear and nonlinear models 
in Section \ref{SimulationRobust}. 
We end the paper with some concluding remarks. 

\section{The T-Robust and T-Trend smoothing problems}
\label{StudentT-KF} 

\begin{figure*} \label{GLT-KF}
  \begin{center} 
  \begin{tabular}{ccc} 
  \includegraphics[scale=0.39]{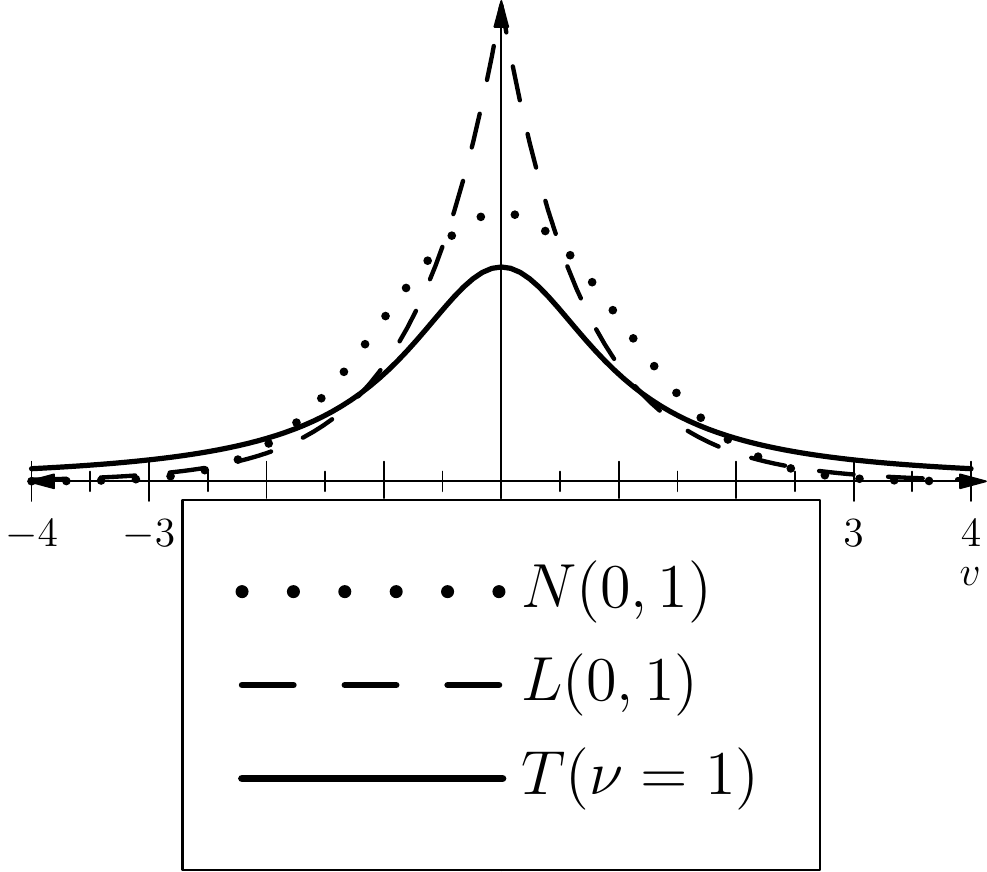}
\hspace{.1in}
\includegraphics[scale=0.39]{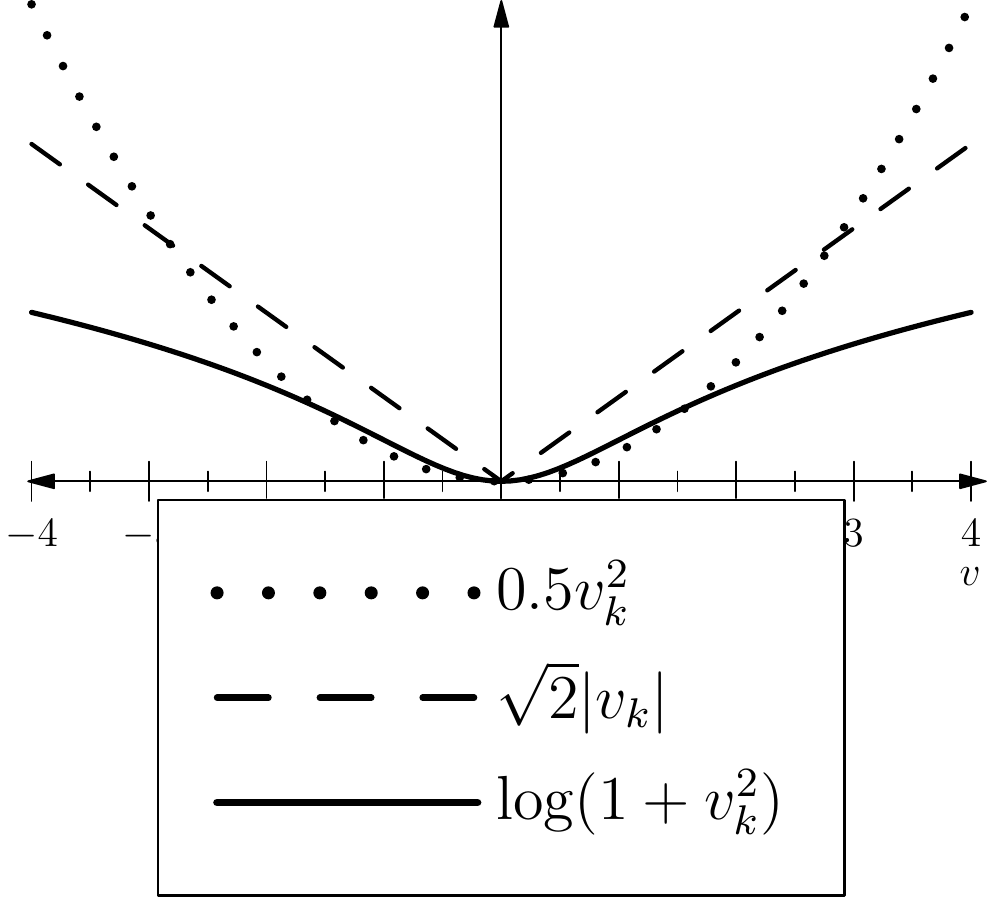}
\hspace{.1in}
\includegraphics[scale=0.39]{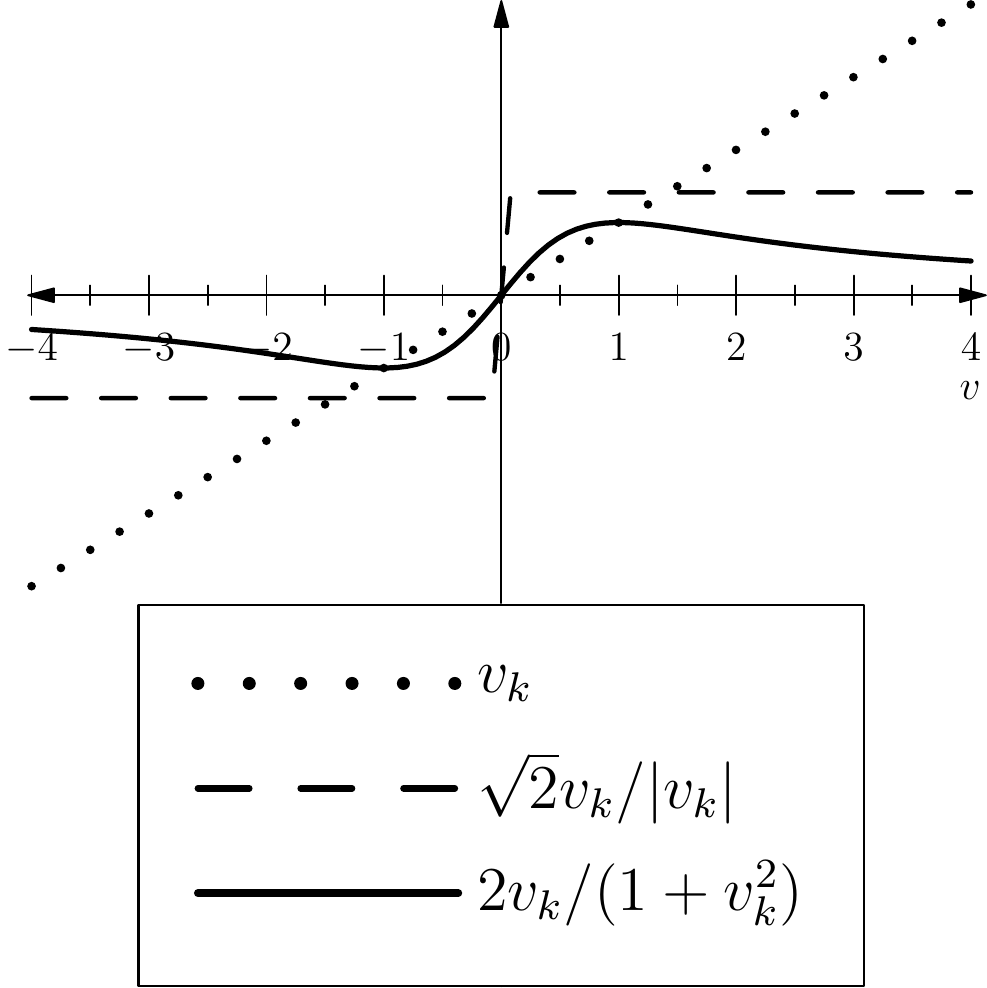}
    \end{tabular}
    \caption{Gaussian, Laplace, and Student's t Densities, Corresponding Negative Log Likelihoods, and Influence Functions (for scalar $v_k$).}
     \end{center}
\end{figure*}

For a vector \(u \in \mB{R}^n\)
and any positive definite matrix \(M \in \mB{R}^{n\times n}\), let
\(\|u\|_M := \sqrt{u^\R{T}Mu}\).
We use the following generalization of the Student's 
t-distribution: 
\begin{eqnarray}
\label{StudentDensity}
\B{p}(v_k|\mu)
&=&
 \frac{\Gamma (\frac{s + m}{2})}
{\Gamma(\frac{s}{2})\det[\pi s R]^{1/2}}
\left(1 + \frac{\|v_k - \mu\|_{R^{-1}}^2}{s}\right)^{\frac{-(s + m)}{2}}
\end{eqnarray}
where $\mu$ is the mean, $s$ is the degrees of freedom,
$m$ is the dimension of the vector $v_k$, 
and $R$ is a positive definite matrix.
A comparison of this distribution
with the Gaussian and Laplacian distribution 
appears in Figure \ref{GLT-KF}. Note that the Student's t-distribution
has much heavier tails than the others, and that its influence
function is re-descending, see \citep{Mar} for a discussion of 
influence functions. This means that as we pull a 
measurement further and further away, its `influence'
decreases to 0, so it is eventually ignored by the model.  
Note also that the $\ell_1$-Laplace is peaked at 0, while the 
Student's t-distribution is not, and so a Student's t-fit 
will not in general drive residuals to be exactly $0$.\\ 
We use the following general model for the underlying dynamics:
for $k = 1 , \ldots , N$
\begin{equation}
\label{NonlinearStudentModel}
\begin{array}{ccc}
	x_{k} & = & g_k(x_{k-1}) + w_k
	\\
	z_k   & = & h_k(z_k) + v_k
\end{array}
\end{equation}
with initial condition $g_1(x_0) = g_0 + w_1$, with $g_0$ a known constant, and
where $g_k: \mB{R}^n \rightarrow \mB{R}^n$ are known smooth process functions,
and $h_k: \mB{R}^n \rightarrow \mB{R}^{m(k)}$ are known smooth
measurement functions. \\
For the T-Robust smoother, we assume that 
the vector $w_k \in \mB{R}^n$ is zero-mean Gaussian noise
of known covariance $Q_k \in \mB{R}^{n\times n}$,
and the vector $v_k \in \mB{R}^{m(k)}$
is zero-mean Student's t measurement noise \eqref{StudentDensity}
of known covariance $R_k \in \mB{R}^{m(k) \times m(k)}$ and degrees of freedom $s$.\\
For the T-Trend smoother, the roles are interchanged, i.e. $w_k$ is Student's t noise
while $v_k$ is Gaussian. In both cases, we assume that the vectors $ \{ w_k \} \cup \{ v_k \} $
are all mutually independent.\\
In the next sections, we design methods to find the MAP estimates of $\{x_k\}$
for both formulations. 

\section{T-Robust smoother}
\label{ML}

Given a sequence of column vectors $\{ u_k \}$
and matrices $ \{ T_k \}$ we use the notation
\[
\R{vec} ( \{ u_k \} )
=
\begin{bmatrix}
u_1 \\ u_2  \\ \vdots \\ u_N
\end{bmatrix}
\; , \;
\R{diag} ( \{ T_k \} )
=
\begin{bmatrix}
T_1    & 0      & \cdots & 0 \\
0      & T_2    & \ddots & \vdots \\
\vdots & \ddots & \ddots & 0 \\
0      & \cdots & 0      & T_N
\end{bmatrix} .
\]
We also make the following definitions:
\[
\begin{array}{rcl}
R       & = & \R{diag} ( \{ R_k \} ) 
\\
Q       & = & \R{diag} ( \{ Q_k \} ) 
\\
x       & = & \R{vec} ( \{ x_k \} )
\end{array}
\; , \;
\begin{array}{rcl}
w( x )    & = & \R{vec} ( \{ x_k - g_k ( x_{k-1} ) \} )
\\
v ( x )  & = & \R{vec} ( \{ z_k - h_k( x_k ) \} ) . 
\end{array}
\]
Maximizing the likelihood for the model (\ref{NonlinearStudentModel}) 
is equivalent to minimizing the associated negative log likelihood 
\[
-\log \B{p}(\{\nu_k\}, \{w_k\}) 
=
-\log\B{p}(\{\nu_k\}) - \log\B{p}(\{w_k\})
\]
Dropping the terms that do not depend on \(\{x_k\}\), 
the objective corresponding to T-Robust is
\begin{eqnarray}
\label{fullObjectiveRobust}
\frac{1}{2}\sum^N_{k=1}(s_k + m_k)\log
\left[1 + \frac{\|v_k\|_{R_{k}^{-1}}^2}{s_k}\right]
+
\|w_k\|_{Q_k^{-1}}^2 ,
\end{eqnarray}
where $s_k$'s are degrees of freedom parameters
associated with measurement noise,
and $m_k$ are the dimensions of the $k$th observation. 

A first-order accurate
affine approximation to our model 
with respect to direction 
$d = \R{vec}\{d_k\}$ 
near a fixed state sequence $x$ is given by 
\[
\begin{array}{lll}
\tilde w( x;d )    
& = & 
\R{vec} ( \{ x_k - g_k(x_{k-1}) - g_k^{(1)} (x_{k-1}) d_k \} ),
\\
\tilde v ( x;d  ) 
& = & 
\R{vec} ( \{ z_k -  h_k(x_k) - h_k^{(1)} (x_k) d_k \} ). 
\end{array}
\]
Set $Q_{N+1} = I_n$ and $g_{N+1} ( x_N ) = 0$ 
(where $I_n$ is the $n \times n$ identity matrix)
so that the formulas are also valid for $k = N+1$.

We minimize the nonlinear nonconvex objective in (\ref{fullObjectiveRobust}) 
by iteratively solving quadratic programming (QP) subproblems of the form: 
\begin{equation}
\label{QuadraticSubproblemOne}
\begin{array}{lll}
\mbox{min} 
	&\frac{1}{2}d^\R{T}Cd + a^\R{T}d  
	\quad \mbox{w.r.t} \;  d \in \mB{R}^{nN} , 
\end{array}
\end{equation}
where $a$ is the gradient of objective (4) with respect
to $x$ and $C$ has the form 
\begin{equation}
\label{hessianApprox}
C 
= 
\begin{bmatrix} 
C_1 & A_2^\R{T} & 0 & \\
A_2 & C_2 & A_3^\R{T} & 0 \\
0 & \ddots & \ddots& \ddots & \\
& 0 & A_N & C_N
\end{bmatrix} ,
\end{equation}
with $A_k \in \mB{R}^{n\times n}$ and 
$C_k \in \mB{R}^{n\times n}$ defined as follows: 
\begin{eqnarray}
\nonumber 
A_k 
&=& 
-Q_k^{-1}g^{(1)}_{k}\; , \;\\
\nonumber 
C_k 
&=& 
Q_k^{-1} + (g^{(1)}_{k+1})^\R{T}Q^{-1}_{k+1}g^{(1)}_{k+1} + H_k\; ,\\
\nonumber
H_k
&=& 
\frac{(h_k^{(1)})^\R{T}R_k^{-1}h_k^{(1)}}
{(s_k + \|v_k\|_{R^{-1}_k}^2)/(s_k + m_k)}\; .
\end{eqnarray}
The solutions to the subproblem \eqref{QuadraticSubproblemOne} have the form 
$d = -C^{-1}a$, 
and can be found in an efficient 
and numerically stable manner in $O(n^3N)$ steps, 
since $C$ is tridiagonal and positive definite (see \citep{Bell2008}).

\section{T-Trend Smoother}
\label{T-Trend}
The objective corresponding to T-Trend is
\begin{equation}
\label{fullObjectiveTrend}
\frac{1}{2}\sum^N_{k=1}(r_k + n)
\log \left[1 + \frac{\|w_k\|_{Q_{k}^{-1}}^2}{r_k}\right]
+
\|v_k\|_{R_k^{-1}}^2 ,
\end{equation}
where $r_k$ are degrees of freedom parameters
associated with process noise,
and $n$ is the dimension of each state $x_k$. 
A first-order accurate
affine approximation to our model 
with respect to direction 
$d = \R{vec}\{d_k\}$ 
near a fixed state sequence $x$ is as follows: 
\[
\begin{array}{lll}
\tilde w( x;d )    
& = & 
\R{vec} ( \{ x_k - g_k(x_{-1}) - g_k^{(1)} (x_{k-1}) d_k \} ),
\\
\tilde v ( x;d  ) 
& = & 
\R{vec} ( \{ z_k -  h_k(x_k) - h_k^{(1)} (x_k) d_k \} ). 
\end{array}
\]
As before, we set $Q_{N+1} = I_n$ and $g_{N+1} ( x_N ) = 0$ 
(where $I_n$ is the $n \times n$ identity matrix)
so that the formula is also valid for $k = N+1$.

We minimize the nonlinear objective in (\ref{fullObjectiveTrend}) 
by iteratively solving quadratic programming (QP) subproblems of the form 
\begin{equation}
\label{TrendQuadraticSubproblemOne}
\begin{array}{lll}
\mbox{min} 
	&\frac{1}{2}d^\R{T}Cd + a^\R{T}d  
	\quad \mbox{w.r.t} \;  d \in \mB{R}^{nN} , 
\end{array}
\end{equation}
where $a$ is the gradient of objective (4) with respect
to $x$ and $C$ again has form ~\eqref{hessianApprox}, 
but now 
with $A_k \in \mB{R}^{n\times n}$ and 
$C_k \in \mB{R}^{n\times n}$ defined as follows: 
\begin{eqnarray}
\nonumber 
A_k 
&=& 
-\frac{(r_k + n)Q_k^{-1}g^{(1)}_{k}}{r_k + \|w_{k+1}\|_{Q^{-1}_k}^2},\\
\nonumber 
C_k 
&=& 
Q_k^{-1} + (h_k^{(1)})^\R{T}R_k^{-1}h_k^{(1)} + H_k,\\
\label{TrendH}
H_k
&=& 
\frac{(g^{(1)}_{k+1})^\R{T}Q^{-1}_{k+1}g^{(1)}_{k+1}}
{(r_k + \|w_k\|_{Q^{-1}_k}^2)/(r_k + n)}.
\end{eqnarray}
The solutions to the subproblem \eqref{TrendQuadraticSubproblemOne} 
again have the form 
$d = -C^{-1}a$, where $C$ is tridiagonal and positive definite, 
so that they still can be found in an efficient 
and numerically stable manner in $O(n^3N)$ steps, 
see \citep{Bell2008}.  

\section{Algorithm and Global Convergence} 
\label{Algorithm}

When models $g_k$ and $h_k$ are all linear, 
we can compare the algorithmic scheme proposed
in the previous sections with the method
in \citep{Fahr1998}.
The method in \citep{Fahr1998} proposes 
using the Fisher information matrix in place
of the matrix $C$ above. 
When the densities for $w_k$ and $v_k$
are Gaussian, this is equivalent to the Gauss-Newton
method. However, in the Student's t-case, 
the Fisher information matrix may be indefinite. \\
When this occurs, \citep{Fahr1998} propose 
to use the expectation 
of the Fisher information matrix, i.e. the 
Fisher scoring method. In this approach 
the expected value of $C$ replaces 
the terms $\|w_k\|_2^2$ or $\|v_k\|_2^2$
in the denominators of $H_k$ and $A_k$ with
their expectations, which are only functions 
of $s_k$ and $r_k$, the degrees of freedom. 
The crucial difference here is that in practice,
we want to pass the information about 
which $\|w_k\|$ and $\|v_k\|$ are large to the algorithm,
so that it can curtail their contribution to the 
model updates. So while the Fisher information
matrix is too unstable (can become indefinite), 
the expected Fisher information is insensitive
to the magnitude variations the algorithm should incorporate
as it proceeds. \\
For these reasons, we propose a Gauss-Newton method 
which uses the relative size
information of the residuals to find the 
directions of descent, 
and provide a proof of convergence 
for the application of this method to solve
(\ref{fullObjectiveRobust}).
This objective takes the form \(K= \rho \circ F\),
with the convex function \( \rho \) and the smooth function \(F\)
given by
\begin{eqnarray}
\label{rho-T-Robust}
\rho \left( \begin{array}{c} c \\ u \end{array} \right)
& = & 
c
+
\frac{1}{2}\|u\|_{A^{-1}}^2
\\
\label{F-T-Robust}
F(x)
& = &
\left( \begin{array}{c}  f(x)\\ r(x) \end{array} \right)
\\
f(x) 
&=&
 \frac{1}{2}
\sum_{k=1}^N(b_k + l_k) 
\log
\left[
1 + \frac{\|p_k(x)\|_{B_k^{-1}}^2}{b_k}
\right] ,
\end{eqnarray}
where $r(x)$ and $p(x)$ are smooth functions of $x$, $A$ and $\{B_k\}$ 
are known positive definite matrices, $b_k, l_k$ are fixed constants, 
and $c \in \mB{R}$.
This structure covers both T-Robust and T-Trend, 
where for T-Robust $p_k(x) = v_k(x)$, $r(x) = w(x)$, $A = \R{diag}[\{Q_k\}]$ 
and $B_k = R_k$, and for T-Trend $p_k(x) = w_k(x)$, $r(x) = v(x)$, $A = \R{diag}[\{R_k\}]$ and $B_k = Q)k$. Note that the range of $f$ is $\B{R_+}$. \\
The objective $K(x) = \rho \circ F(x)$ is convex composite, 
since $\rho$ is convex and $F(x)$ is smooth. Our 
approach exploits this structure by iteratively 
linearizing $F$ about the iterates $x^k$ and solving
\begin{equation}
\label{DirectionFindingSubproblem}
\begin{array}{lll}
&\displaystyle \min_{d\in \mB{R}^{nN}}& 
\rho(F(x^k) + F^{(1)}(x^k)d) .
\end{array}
\end{equation}
Rather than requiring interior point methods 
to solve~\eqref{DirectionFindingSubproblem}
as in \citep{AravkinL12011}, a single block-tridiagonal
solve of the system~\eqref{QuadraticSubproblemOne}
yields descent direction $d$ for the objective $K(x)$. 
While the details of solving the direction finding subproblem 
are always problem specific, a general convergence theory
for convex-composite methods can be derived to 
establishes the overall convergence to a stationary point 
of $K(x)$. The theory required generalizes 
that of \citep{AravkinL12011}, and includes both 
T-Robust and T-Trend formulations. We present
the theory and remark on its particular application
to the problems of interest. Please see \citep[Theorem 4.5.2, Corollary 4.5.3]{AravkinThesis2010}
for the proofs. \\
Recall the first-order necessary condition for optimality in the convex composite problem minimize $K(x)$ is
\[
0 \in \partial K(x) = \partial \rho \left( F(x) \right) F^{(1)} (x)
\]
where \(\partial K (x)\) is the
generalized subdifferential of \(K\) at \(x\) \cite{RTRW}
and \(\partial \rho \left( F(x) \right) \) is the convex subdifferential
of \( \rho \) at \(F(x)\) \cite{Rock70}.
Elementary convex analysis gives us the equivalence
\[
0 \in \partial K(x)
\quad \Leftrightarrow \quad
K(x) = \inf_d \rho \left( \; F(x) + F^{(1)} (x) d \; \right) \; .
\]
For both T-Robust and T-Trend, 
it is desirable to modify the objective in \eqref{DirectionFindingSubproblem}
by including curvature information.
We therefore define the difference function
\begin{equation}
\label{extendedDelta}
\Delta( x, H; d) = \rho \left( \; F(x) + F^{(1)} (x) d \; \right) 
+
\frac{1}{2}d^\R{T}Hd
 - K(x) \; ,
\end{equation}
where $H = H(x)$ is positive semidefinite and varies continuously 
with $x$, 
and the minimum of $\Delta(x, H; d)$ with respect to direction $d$
\begin{equation}
\label{extendedInf}
\Delta^* ( x, H)      = \inf_d \Delta (x, H ; \; d) \; .
\end{equation}
Since $\frac{1}{2}d^\R{T}Hd$ is differentiable 
at the origin for any $H$, we have \(\Delta^* ( x, H) = 0\) 
if and only if \(0 \in \partial K(x)\) \cite{Burke85}.  \\
Given \(\eta \in (0,1)\), we define a set of search directions at \(x\) by
\begin{equation}
\label{extendedDSet}
D( x, H , \eta ) = \set{d}{ \Delta(x , H; d) \le \eta \Delta^* ( x, H) }  \; .
\end{equation}
Note that if there is a \(d \in D( x, H,\eta)\) such that
\(\Delta( x, H ; \; d) \ge - \eta \varepsilon \), then
\( \Delta^* ( x, H) \ge - \varepsilon\).
These ideas motivate the following
algorithm \cite{Burke85}.

\begin{algor}
\label{GaussNewtonAlgorithm}
{\it Gauss-Newton Algorithm.} 
\begin{enumerate}
\item
Given
\( x^0 \in \mB{R}^{Nn} \) an initial estimate of state sequence,
\(H_0\in\mB{S}^n_+\) the initial curvature information,
\(\varepsilon \ge 0 \) an overall termination criteria,
\( \eta \in (0,1) \) a termination criteria for subproblem,
\( \beta \in (0, 1) \) a line search rejection criteria,
\( \gamma \in (0,1) \) a line search step size factor.
Set the iteration counter \( \nu = 0 \).
\item (Gauss-Newton Step)
\label{GaussNewtonStep}
Find \(d^\nu \) in the set \( D(x^\nu,H_\nu , \eta ) \)
in \ref{extendedDSet}.
Set \(\Delta_\nu = \Delta( x^\nu , H_\nu; d^\nu) \) 
in \ref{extendedDelta} and
{\it Terminate} if \( \Delta_\nu \ge - \varepsilon \).
\item (Line Search) Set
\[
\begin{array}{lll}
t_\nu &=& \max  \gamma^i \\
	 &\text{s.t.}&i \in \{ 0, 1, 2, \cdots \} \; \mbox{ and }
	\\
	&\text{s.t.}&\rho \left( F( x^\nu + \gamma^i d^\nu ) \right)
		\le \rho \left( F( x^\nu ) \right)
                     + \beta \gamma^i \Delta_\nu.
\end{array}
\]
\item (Iterate)
Set \(x^{\nu+1} = x^\nu + t_\nu d^\nu \),
select \(H_{\nu+1}\in\mB{S}^n_+\) and goto Step~\ref{GaussNewtonStep}.
\end{enumerate}
\end{algor}
\begin{theorem}
\label{GlobalConvergenceTheorem}
Let $K(x) = \rho \circ F(x)$, with \( \rho \) be convex and coercive on its domain, \(F\) continuously
differentiable, and further assume that \(x^0\in\mB{R}^{Nn}\) is such that
\(F^{(1)}\) is uniformly continuous on the set
\( \overline{\R{co}}\left(\set{x}{K(x)\le K(x^0)}\right) \).
Fix \(x^0 \in \mB{R}^{Nn} \), define
\[
\Lambda = \set{u}{ \rho (u) \le K(x^0)}\;.
\]
Suppose that \(F^{-1}(\Lambda) = \set{x}{F(x)\in\Lambda}\)
is bounded, and either of the following assumptions hold:
\begin{align}
&0 \leq \lambda_{\min} \leq \mathrm{eig}(H^\nu)\label{AssumptionOne}\\
&\R{Null} \left( F^{(1)}(x) \right)
 = \{ 0 \} \quad \forall \; x \in F^{-1}(\Lambda)\label{AssumptionTwo}\; .
\end{align}
If \(\{ x^\nu \}\) is a sequence generated by Algorithm~\ref{GaussNewtonAlgorithm}
with initial point \(x^0\) and \(\varepsilon = 0\), then \(\{ x^\nu \}\) and
\(\{d^\nu\}\) are bounded and either
the algorithm terminates finitely at a point \( x^\nu \) with
\( 0 \in \partial K( x^\nu ) \), or \( \Delta_\nu \rightarrow 0 \)
as \( \nu \rightarrow \infty \),
and every cluster point
\(\bar{x}\) of the sequence \(\{x^\nu\}\)
satisfies \(0 \in \partial K( \bar{x} )\).
If \(\{ (x^\nu,H_\nu) \}\) is a sequence generated by the 
Gauss-Newton algorithm above
with initial point \((x^0,H_0)\) and \(\varepsilon = 0\).
If the sequence \(\{H_\nu\}\) remains bounded, then \(\{ x^\nu \}\) and
\(\{d^\nu\}\) are bounded and either
the algorithm terminates finitely at a point \( x^\nu \) with
\( 0 \in \partial K( x^\nu ) \), or \( \Delta_\nu \rightarrow 0 \)
as \( \nu \rightarrow \infty \).
Moreover, every cluster point
\(\bar{x}\) of the sequence \(\{x^\nu\}\)
satisfies \(0 \in \partial K( \bar{x} )\).
\end{theorem}

\begin{remark}
Both T-Robust and T-Trend can be solved by
Algorithm \ref{GaussNewtonAlgorithm}, with the objective
function $K(x)$ as above. Note that $\rho$ above
is coercive on the range of $F$. \\
For T-Robust, \eqref{AssumptionTwo} always holds, since 
of $F^{(1)}$ contains a block-bidiagonal matrix with identities on the diagonal. 
For T-Trend, \eqref{AssumptionOne} holds if all \(g_k^{(1)}(x_k)\) are nonsingular for all 
\(x\in F^{-1}(\Lambda)\); see~\eqref{TrendH}. 
\end{remark}

\section{Numerical Experiments}
\label{SimulationRobust} 

\subsection{T-Robust Smoother} 

\begin{table}
\small
\caption{
\label{SimulationResults}
Median MSE over 1000 runs and 
intervals containing 95\% of MSE results}.
\begin{center}
\begin{tabular}{|c|c|c|c|c|}\hline
{\bf Outlier }
    & {p}
    & {\B {KS MSE}}
    & {\B {RKS MSE}} 
    & {\B {TKS MSE}}
\\ \hline
Nom.
 & ---
    &.04(.02, .1)
    &.04(.01, .1)
    & .04(.01, .09) 
\\ \hline
$\B{N}(0, 10)$
& .1
    &.17(.05, .55)
    &.05(.02, .13)
    &.04(.02, .11) 
\\ \hline
$\B{N}(0, 100)$
& .1
    &1.3(.30, 5.0)
    &.05(.02, .14)
    &.04(.02, .11) 
\\ \hline 
$\B{U}(-10, 10)$
& .1
    &.47(.12, 1.5)
    &.05(.02, .13)
    &.04(.02, .10) 
\\ \hline 
$\B{N}(0, 10)$
& .2
    &.32(.11, .95)
    &.06(.02, .19)
    &.05(.02, .16) 
\\ \hline
$\B{N}(0, 100)$
& .2
    &2.9(.94, 8.5)
    &.07(.02, .22)
    &.05(.02, .14) 
\\ \hline 
$\B{U}(-10, 10)$
& .2
    &1.1(.36, 3.0)
    &.07(.03, .26)
    &.05(.02, .13) 
\\ \hline 
$\B{N}(0, 10)$
& .5
   &.74(.29, 1.9)
    &.13(.05, .49)
    &.10(.04, .45) 
\\ \hline
$\B{N}(0, 100)$
& .5
    &7.7(2.9, 18)
    &.21(.06, 1.6)
    &.09(.03, .44) 
\\ \hline 
$\B{U}(-10, 10)$
& .5
    &2.6(1.0, 5.8)
    &.20(.06, 1.4)
    &.10(.03, .44) 
\\ \hline 
\end{tabular}
\end{center}
\end{table}

\subsubsection{Linear Example}

In this section we compare the new T-robust 
smoother with the  $L_2$-Kalman smoother~\citep{Bell2008} 
and with the $\ell_1$-Laplace robust smoother~\citep{AravkinL12011},
both implemented in~\citep{ckbs}. 
The {\it ground truth}
for this simulated example is
\[
x(t) = \begin{bmatrix} -\cos(t) & -\sin(t)\end{bmatrix}^\R{T} \; .
\]
The time between measurements is a constant $\Delta t$.
We model the two components of the state as integral and two-fold integral of the 
same white noise, so that 
$$
g_k ( x_{k-1} ) =
\begin{bmatrix}
	1        & 0
	\\
	\Delta t & 1
\end{bmatrix} x_{k-1}
\; ,
\qquad 
Q_k =
\begin{bmatrix}
	\Delta t   & \Delta t^2 / 2
	\\
	\Delta t^2 / 2 & \Delta t^3 / 3
\end{bmatrix}\;.
$$
The measurement model for the conditional mean of measurement $z_k$ given state $x_k$ is
defined by
$$
h_k( x_k ) =  \begin{bmatrix} 0 & 1 \end{bmatrix} x_k = x_{2,k} \;,
\qquad R_k = \sigma^2\;,
$$
where \( x_{2,k} \) denotes the second component of \( x_k \),
$\sigma^2 = 0.25$ for all experiments, and the degrees of 
freedom parameter $k$ was set to 4 for the Student's t methods.\\
The measurements \( \{z_k \} \) were generated 
as a sample from
\[
\B{z_k} = x_2 ( t_k ) + v_k, \quad k=1,\ldots,100, \quad t_k=0.04\pi \times k
\]
where the measurement noise $v_k$ was
generated according to the following schemes.
\begin{enumerate}
\item
(Nominal): $v_k \sim \B{N} ( 0 , 0.25 )$ 
\item
(Contaminating Normal) $v_k \sim (1 - p ) \B{N} (0, 0.25 ) + p \B{N} ( 0 , \phi )$,
for $p \in \{ 0.1 , 0.2, 0.5 \}$
and $\phi \in \{10, 100 \}$.
\item 
(Contaminating Uniform) Same as above, but with $\B{U}[-10, 10]$ replacing normal
contamination, and $p = 0.1, 0.2, 0.5$. 
\end{enumerate}

The results for our simulated fitting are presented in
Table~\ref{SimulationResults}.
Each experiment was performed 1000 times, and we provide the
median Mean Squared Error (MSE) value and a quantile interval containing
95\% of the results. The MSE is defined by
\begin{equation}
\label{MSEeq}
	\frac{1}{N} \sum_{k=1}^N
		[ x_1 ( t_k ) - \hat{x}_{1,k} ]^2
		+
		[ x_2 ( t_k ) - \hat{x}_{2,k} ]^2 , 
\end{equation}
where \( \{ \hat{x}_k \} \) is the corresponding estimating sequence.

Note that both of the smoothers perform as well as the (optimal) 
$L_2$-smoother at nominal conditions, and that both continue to perform
at that same level for a variety of outlier generating scenarios. 
The T-smoother always performs at least as well 
as the $\ell_1$-smoother, and it gains an advantage
when either the probability of contamination is high, 
or the contamination is uniform. This is likely due to 
the re-descending influence function of the Student's t-distribution --- 
the smoother effectively throws out bad points rather than simply 
decreasing their impact to a certain threshold, as is the case for the 
$\ell_1$-smoother. 

\subsubsection{Nonlinear Example}
\label{NumexpNonlinear}
\begin{figure}
\begin{center}
  {\includegraphics[scale=0.4]{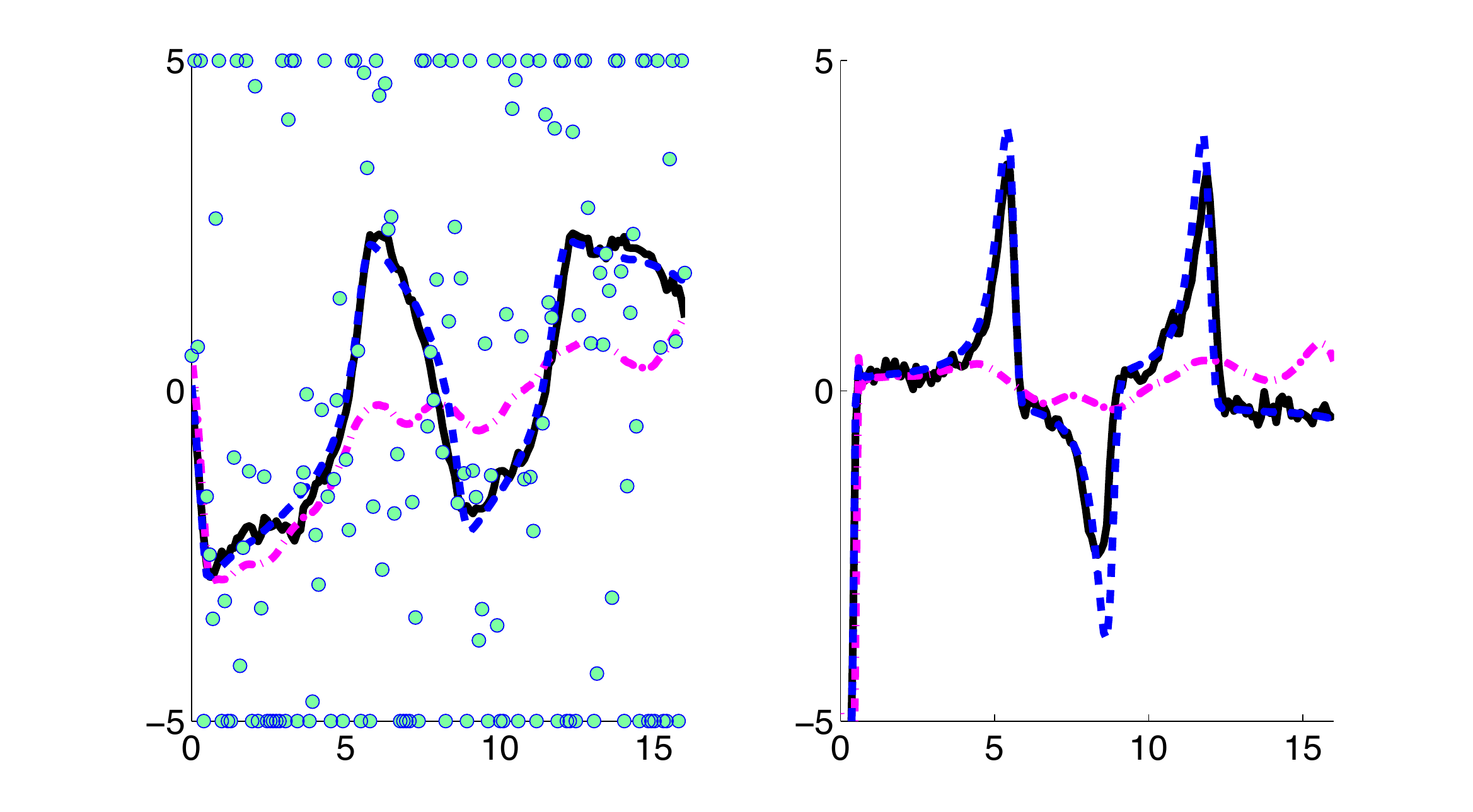}}
\end{center}
\caption{Smoother fits for X-component (left) and Y-component (right) of 
the Van Der Pol oscillator, with 
{\bf 70\% outliers $N(0, 100)$}. 
Black solid line is truth, magenta dash-dot is the $\ell_1$
smoother result, and blue dashed line is T-robust. Measurements on X-component
are shown as dots, with outliers outside the range $[-5, 5]$ plotted on top and bottom axes. \label{VanDerPolFig}}
\end{figure}

In this section, we present results for the 
Van Der Pol oscillator (VDP), described in detail 
in ~\citep{AravkinL12011}. The VDP oscillator is a 
coupled nonlinear ODE 
\[
\dot{X}_1 (t) = X_2 (t)
\hspace{.5cm} \mbox{and} \hspace{.5cm}
\dot{X}_2 (t) = \mu [ 1 - X_1 (t)^2 ] X_2 (t) - X_1 (t)
\; .
\]
The process model here is the Euler approximation for $X(t_k)$
given $X(t_{-1})$:
\[
g_k ( x_{k-1} ) = \left( \begin{array}{cc}
	x_{1,k-1} + x_{2,k-1} \Delta t
	\\
	x_{2,k-1} + \{ \mu [ 1 - x_{1,k}^2 ] x_{2,k} - x_{1,k} \} \Delta t
\end{array} \right) \; .
\]
For this simulation, the {\it ground truth} is obtained from
a stochastic Euler approximation
of the VDP.
To be specific,
with \( \mu = 2 \), \( N = 164 \) and \( \Delta t = 16 / N \),
the ground truth state vector \( x_k \) at time \( t_k = k \Delta t \)
is given by \( x_0 = ( 0 , -0.5 )^\R{T} \) and
for \( k = 1, \ldots , N \),
$x_k = g_k ( x_{k-1} ) + w_k$,
where $\{ w_k \}$ is a realization of
independent Gaussian noise with variance $0.01$.\\
The $\ell_1$-Laplace smoother was shown to have superior
performance to the Gaussian nonlinear smoother in 
\citep{AravkinL12011}, both implemented in ~\citep{ckbs}.
We compared the performance of the nonlinear T-robust and nonlinear
 $\ell_1$-Laplace smoothers, and found that T-robust gains
 an advantage in the extreme cases of 70\% outliers (see Figure~\ref{VanDerPolFig}), and otherwise is hard to distinguish from the $\ell_1$-Laplace for 40\% or 
 fewer outliers.

\subsection{T-Trend Smoother} 
\label{NumexpT-Trend}

\begin{figure}
\begin{center}
  {\includegraphics[scale=0.32,angle=-90]{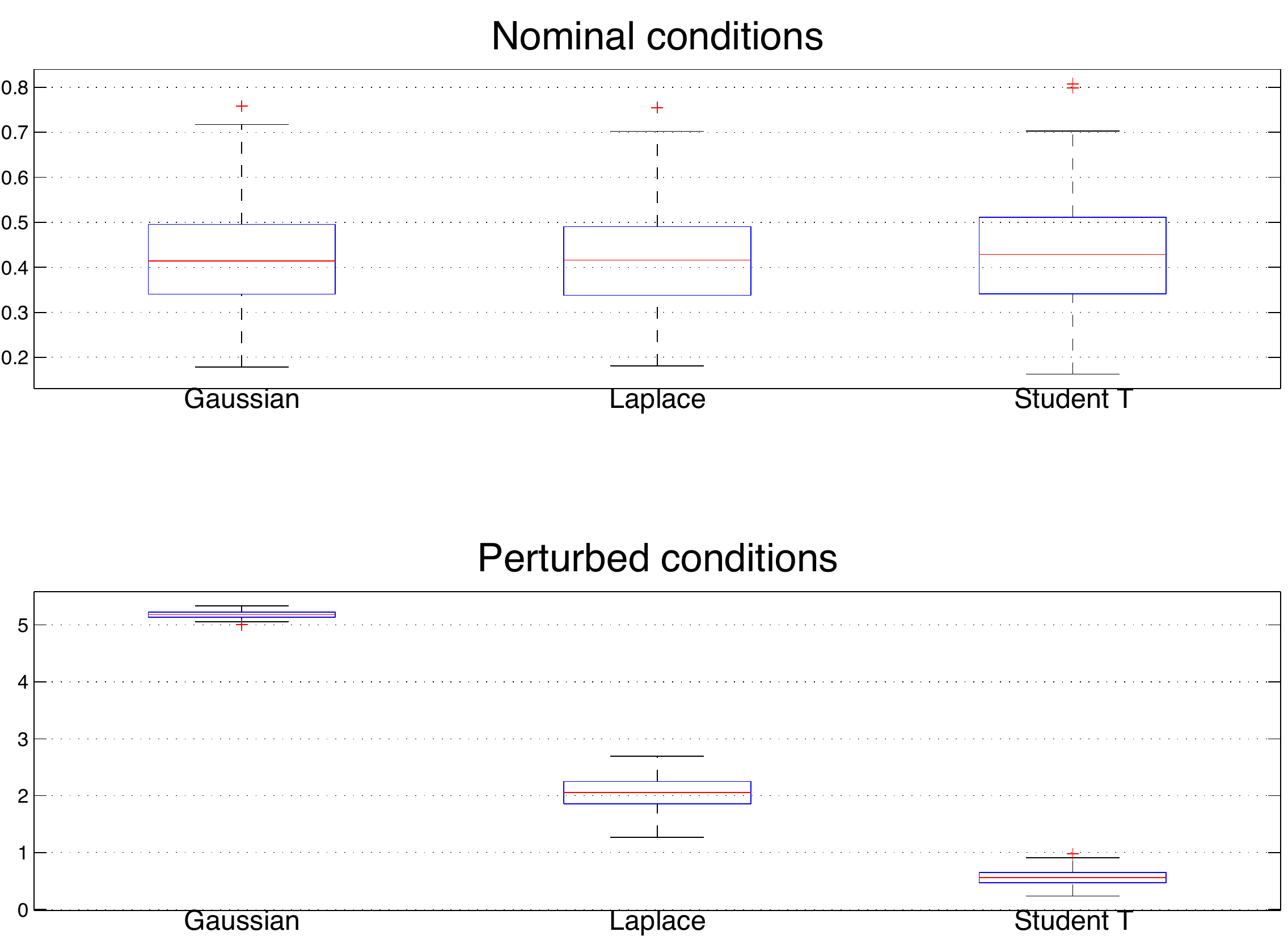}}
\end{center}
\caption{T-Trend Smoother: Monte Carlo simulation. Boxplot of errors obtained using Gaussian, Laplace and Student's T Kalman smoother
under nominal (top) and perturbed (bottom) conditions. \label{jumpSmoother1}
}
\end{figure}

\begin{figure}
\begin{center}
  {\includegraphics[scale=0.32,angle=-90]{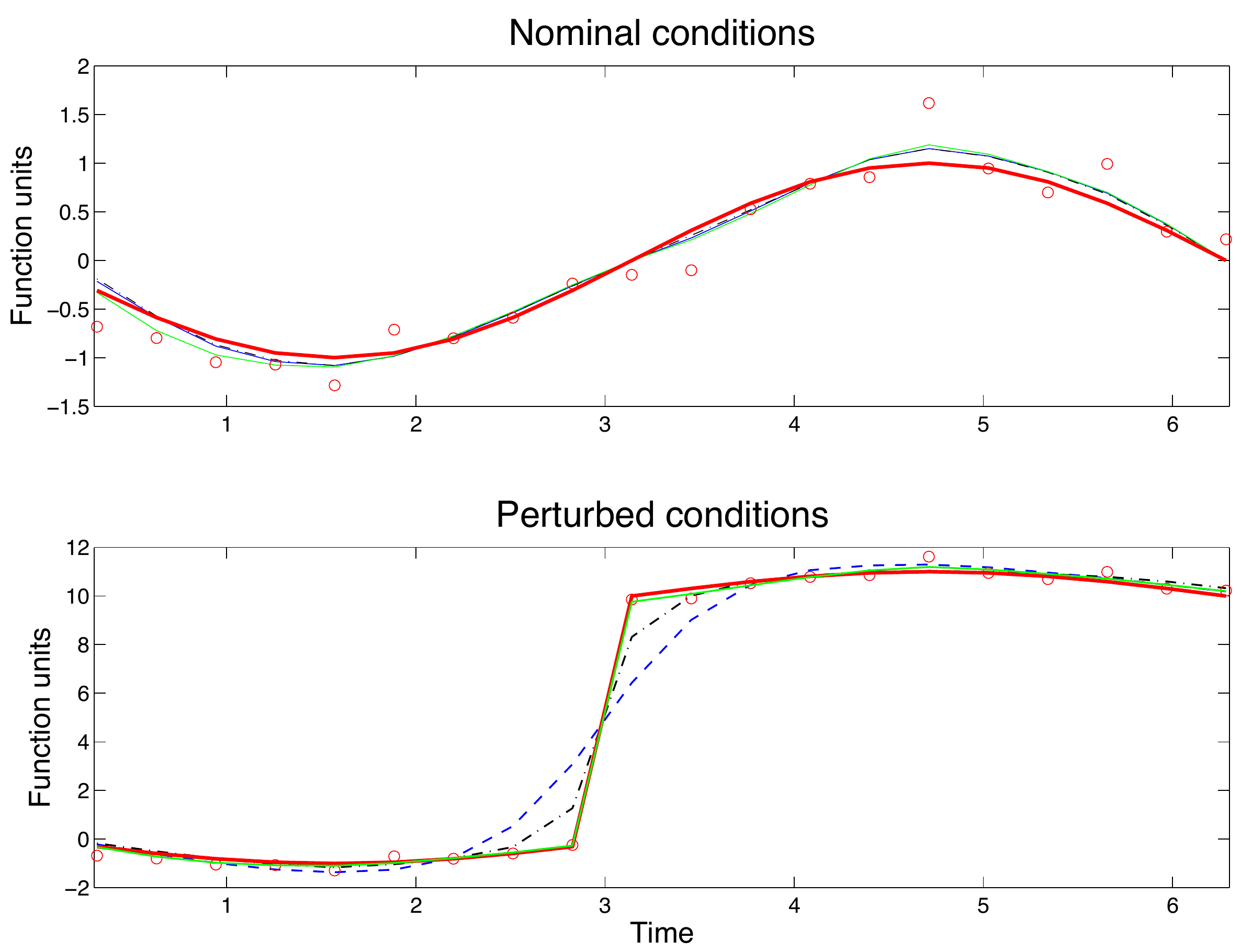}}
\end{center}
\caption{T-Trend Smoother: results from a Monte Carlo run under nominal (top)
and perturbed (bottom) conditions using $L_2$ (dashed),
$L_1$ (dashdot) and $T-Trend$ (thin line) smoother. The thick line is the true state.
\label{jumpSmoother2} }
\end{figure}

We present a proof of concept result for the T-Trend smoother,
in particular considering two Monte Carlo studies of 200 runs.
In the first study, the state vector, as well as the process 
and measurement models, are exactly the same as in the linear example
used for the T-Robust smoother in the previous
subsection. At any run,  $x_2$ has to be reconstructed
from 20 measurements corrupted by a white Gaussian noise of variance 0.05 and
collected on $[0,2\pi]$ using a uniform sampling grid.
The top panel of Figure \ref{jumpSmoother1} reports the boxplot of the 200 root-MSE errors, with MSE defined by
	\(\sqrt {\frac{1}{N} \sum_{k=1}^N
		[ x_2 ( t_k ) - \hat{x}_{2,k} ]^2}\), 
obtained using the $L_2$-, $\ell_1$-, and T-Trend Kalman smoothers,
while the top panel of Figure \ref{jumpSmoother2} displays the estimate
obtained in a single run. It is apparent that the performance of the three estimators
is very similar.\\ 
The second experiment is identical to the first one except that
we introduce a `jump' at the middle of the sinusoidal wave. 
The bottom panel of Figure \ref{jumpSmoother1}
reveals the superior performance of the T-Trend smoother
under these perturbed conditions. Further, 
the bottom panel of Figure \ref{jumpSmoother2} shows 
that the estimate achieved by the $L_2$-smoother (dashed-line)
does not follow the jump well (the true state is the solid line). 
The $\ell_1$-smoother (dashdot) does a better job than the $L_2$-smoother, 
but the T-trend smoother outperforms the $\ell_1$-smoother, 
following the jump very closely while still providing a good solution along 
the rest of the path.

\section{Conclusion}

We have described two new nonlinear smoothers,
called T-Robust and T-Trend, which efficiently 
obtain the MAP estimates of the states in a state-space model 
with Student's t measurement and Student's t
process noise, respectively.\\ 
The T-Robust smoother compares favorably to the 
$\ell_1$-Laplace smoother ---  the smoothers are comparable for most 
error scenarios presented, and the T-smoother 
has an advantage for high levels of 
contamination because of the re-descending influence function 
of the Student's t-distribution. 
In addition, although it involves a non-convex objective,
 it is simple to implement using minor modifications
 to the $L_2$ nonlinear smoother.    \\ 
The T-Trend smoother was designed for tracking signals
with potential sudden changes, 
and has many potential applications in 
navigation and financial trend tracking.
It was demonstrated to follow a fast jump in the state better than 
a smoother with a convex penalty on the innovation.  
Just as the T-robust smoother, it can be implemented
with minor modifications to an $L_2$ nonlinear smoother.

\bibliographystyle{plain}
\bibliography{biblio,tks}

\section{Appendix: Full convergence theory}
If \(\{x^\nu\}\) is a sequence generated by the 
Gauss-Newton Algorithm~\ref{GaussNewtonAlgorithm}
with initial point \(x^0\) and \(\varepsilon =0\), then one of the
following must occur:
\begin{enumerate}
\item[(i)]
The algorithm terminates finitely at a point \(x^\nu\) with
\(0\in\partial K(x^\nu)\).
\item[(ii)]
\(\lim_{\nu\in I}\Delta_\nu=0\) for every subsequence  \(I\) for which
the set \( \set{d^\nu}{\nu \in I} \) is bounded.
\item[(iii)]
The sequence \(\| d^\nu \|\) diverges to \(+\infty\).
\end{enumerate}

Moreover, if $\bar x$ is any cluster point 
of a subsequence $I \subset \B{Z_+}$
such that the subsequence \(\{d^{\nu}|\nu \in I\}\)
is bounded, then $0 \in \partial K(\bar x)$. 

{\bf{Proof:}} Assertions (i), (ii), and (iii) 
are a restatement of \cite[Theorem 2.4]{Burke85} 
in our context, where
the sets \(D_\nu\) in \cite[Theorem 2.4]{Burke85} are given by
\(D_\nu = D( x^\nu , \eta ) \). 
The requirement that \( \rho \) be Lipschitz continuous
on the set \(\set{(u)}{ \rho (u)\le K(x^0)}\) 
is an immediate consequence of
the fact that \( \rho \) is coercive its domain, so this set is compact. 
This completes the proof of (i), (ii), and (iii). 
By compactness, the matrices $H$ are uniformly continuous
in $x$ on this set.
Suppose that \(\bar x\)
is a cluster point of a sequence \(I \subset \B{Z_+}\)
for which \(\{d^\nu\}\) is bounded. 
Since $\bar{x}$ is a cluster point of $\{x^{\nu}\}$, 
we can take a convergent subsequence along which 
\(\{d^\nu\}\) are still bounded, and by 
continuity
$H^{\nu}$ converge to $\bar H = H(\bar x)$.  
By Bolzano-Weierstrass, we can then find a subsequence 
\(J \subset I\) and vector $\bar d \in \mB{R}^{Nn}$ 
such that 
\((x^\nu, d^\nu) \rightarrow_J ( \bar{x}, \; \bar{d} )\). 
Fix any other point \(\hat d \in \mB{R}^{Nn}\). 
By construction, we have 
\[
\begin{array}{lll}
\Delta_\nu 
&=& 
\rho\left(F(x^\nu) 
+ 
F^{(1)}(x^\nu)d^\nu\right) 
+
\frac{1}{2}\|d^\nu\|_{H^{\nu}}^2 
-
\rho\left(F(x^\nu)\right)\\
&\leq&
\eta \Delta^*(x^\nu) \\
&\leq&
\eta\left(\rho\left(F(x^\nu) 
+ 
F^{(1)}(x^\nu)\hat d \right) 
+
\frac{1}{2}\|\hat d\|_{H^{\nu}}^2 
-
\rho\left(F(x^\nu)\right)\right)\;.
\end{array}
\]
Taking the limit over $J$ gives 
\[
\begin{array}{lll}
0
& = & 
\rho\left(F(\bar x) 
+ 
F^{(1)}(\bar x)\bar d\right)
+
\frac{1}{2}\|\bar d\|_{\bar H}^2 
-
\rho\left(F(\bar x)\right)\\
&\leq&
\eta\left(\rho\left(F(\bar x) 
+ 
F^{(1)}(\bar x)\hat d \right)
+
\frac{1}{2}\|\hat d\|_{\bar H}^2 
-
\rho\left(F(\bar x)\right)\right)\;.
\end{array}
\]
But $\hat d$ was an arbitrary point in 
\(\mB{R}^{Nn}\), so in particular we must have
$\Delta^*(\bar x) = 0$. 
\vspace{-.55cm}
\begin{flushright}
$\blacksquare$
\end{flushright}

A stronger convergence
result is possible under stronger assumptions on \(F\) and \(F^{(1)}\), 
or on the sequence $H^\nu$. 
Fix \(x^0 \in \mB{R}^{Nn} \), and define
\begin{equation}
\label{LevelSetDef}
\Lambda = \set{u}{ \rho (u) \le K(x^0)}\;, 
\end{equation}
where $\rho$ is as above. Note that $\Lambda$
is compact, since $\rho$ is coercive on its domain.

\begin{corollary}
\label{GlobalConvergenceCorollary}
Suppose that \(F^{-1}(\Lambda) = \set{x}{F(x)\in\Lambda}\)
is bounded, and either of the following assumptions hold:
\begin{align}
&0 \leq \lambda_{\min} \leq \mathrm{eig}(H^\nu)\label{AssumptionOneProof}\\
&\R{Null} \left( F^{(1)}(x) \right)
 = \{ 0 \} \quad \forall \; x \in F^{-1}(\Lambda)\label{AssumptionTwoProof}\; .
\end{align}
If \(\{ x^\nu \}\) is a sequence generated by Algorithm~\ref{GaussNewtonAlgorithm}
with initial point \(x^0\) and \(\varepsilon = 0\), then \(\{ x^\nu \}\) and
\(\{d^\nu\}\) are bounded and either
the algorithm terminates finitely at a point \( x^\nu \) with
\( 0 \in \partial K( x^\nu ) \), or \( \Delta_\nu \rightarrow 0 \)
as \( \nu \rightarrow \infty \),
and every cluster point
\(\bar{x}\) of the sequence \(\{x^\nu\}\)
satisfies \(0 \in \partial K( \bar{x} )\).
\end{corollary}

{\bf{Proof:}} First note that \(F^{-1}(\Lambda)\) is closed since \(F\) is continuous and
 \(\Lambda\) is compact, therefore \(F^{-1}(\Lambda)\) is compact.
 Hence \(\overline{\R{co}}\left( F^{-1}(\Lambda) \right)\) is also
 compact. Therefore, \(F^{(1)}\) is uniformly continuous on
 \(\overline{\R{co}}\left( F^{-1}(\Lambda) \right) \) which implies
 that the hypotheses of Theorem \ref{GlobalConvergenceTheorem} are
 satisfied, and so one of (i)-(iii) must hold. If (i) holds we are done, so we
 will assume that the sequence  \(\{x^\nu\}\) is infinite.  Since \(\{ x^\nu \} \subset F^{-1}(\Lambda)\), this sequence is bounded.
 We now show that the sequence \(\{d^\nu\}\) of search directions is also bounded.\\
 Suppose that \eqref{AssumptionOneProof} holds. 
 For any direction $d^{\nu}$, note that $d^{\nu}$ solves 
\[
\begin{array}{lll}
&\displaystyle \min_d& 
\rho\left(F(x) + F^{(1)}(x)d\right) + \frac{1}{2}\|d\|_{H^\nu}^2\;.
\end{array}
\] 
Therefore we have 
\begin{equation}
\label{SolutionBound}
\rho\left(F(x) + F^{(1)}(x)d^{\nu}\right) 
+ \frac{1}{2}\|d^{\nu}\|_{H^\nu}^2 \leq \rho\left(F(x)\right)
\end{equation}
since we can achieve $\rho\left(F(x)\right)$ with $d = 0$.
 Since $\rho\geq 0$, we must have 
 $\{ \frac{1}{2}\|d^{\nu}\|_{H^\nu}^2\} \leq \rho\left(F(x)\right)$, 
 hence $\{\|d^\nu\|_{H^\nu}\}$ are bounded, and 
 $d^\nu$ are bounded by~\eqref{AssumptionOneProof}. \\
Suppose instead that~\eqref{AssumptionTwoProof} holds.  
 We claim that there exists \( \kappa > 0 \) such that
 \[
 \kappa \| d \| \le \| F^{(1)}(x) d \|
	\quad \forall \; d \in \mB{R}^{Nn}
 	\mbox{ and } x \in F^{-1}(\Lambda)\;.
 \]
Indeed, if this were not the case, then there would exist sequences
\( \{ y^i \} \subset F^{-1}(\Lambda) \) and
\( \{ d^i \} \subset \mB{R}^{Nn} \)
such that \( d^i \neq 0 \) and
\begin{equation*}
\| d^i \| / i > \| F^{(1)}( y^i) d^i \|
	\quad \forall \; i = 1, 2, \dots \;.
\end{equation*}
The set \(F^{-1}(\Lambda)\) is compact, hence there exists
a subsequence \( J \subset \B{Z}_+ \),
vector \( \bar{x} \in F^{-1}(\Lambda) \),
and vector \( \bar{d} \in  \mB{R}^{Nn} \)
with \( \| \bar{d} \| = 1\),
such that \( x^\nu \rightarrow_J \bar{x} \) and
\( d^i / \| d^i \| \rightarrow_J \bar{d} \).
It follows from the inequality above that
\[
\frac{1}{i}\ge \left\| \; F^{(1)} ( x^i ) 
\frac{d^i}{\| d^i \|} \; \right\| \;.
\]
Take the limit with respect to the subsequence \( J \) we obtain
\( 0 \ge \| F^{(1)}( \bar{x} ) \bar{d} \| \).
Thus \( \bar{d} \) is in the kernel of \( F^{(1)}( \bar{x} ) \)
and \( \bar{d} \neq 0 \).
This contradicts~\eqref{AssumptionTwoProof} and thereby proves the claim.
For any direction $d^{\nu}$, 
\eqref{SolutionBound} hods and 
$F(x) + F^{(1)}(x)d^{\nu} \in \Lambda$ since $\frac{1}{2}\|d\|_{H}^2 \geq 0$.
Since \(\Lambda\) is compact, and 
\(\{F(x^\nu),\;F(x^\nu)+F^{(1)}(x^\nu)d^\nu\}\subset\Lambda\) by
construction, there is an \( \alpha > 0 \) such that
\( \| u \| \leq \alpha \) for all \( u \in \Lambda \) and
for \( \nu = 1 , 2 , \ldots , \)
\begin{eqnarray*}
\kappa \| d^\nu \|
& \le  &
\| \; F^{(1)}( x^\nu ) d^\nu \; \|
\\
& \le  &
\| \; F( x^\nu ) + F^{(1)}( x^\nu ) d^\nu \; \|
	+ \| F( x^\nu ) \|
\; \le \; 2 \alpha\;.
\end{eqnarray*}
Hence the sequence \(\{d^\nu\}\) of search directions is bounded.\\
In both cases, Theorem \ref{GlobalConvergenceTheorem} tells us that
\( \Delta_\nu \rightarrow 0 \) as \( \nu \rightarrow \infty \).
The final statement of the corollary follows immediately from the final statement
of Theorem \ref{GlobalConvergenceTheorem}.
\vspace{-.55cm}
\begin{flushright}
$\blacksquare$
\end{flushright}

\end{document}